# Emergence of clonal selection and affinity maturation in an ab initio microscopic model of immunity

Muyoung Heo, Konstantin B. Zeldovich, Eugene I. Shakhnovich

Department of Chemistry and Chemical Biology

Harvard University, 12 Oxford Street, Cambridge, MA 02138

**Abstract**

Mechanisms of immunity, and of the host-pathogen interactions in general are among the most fundamental problems of medicine, ecology, and evolution studies. Here, we present a microscopic, protein-level, sequence-based model of immune system, with explicitly defined interactions between host and pathogen proteins. Simulations of this model show that possible outcomes of the infection (extinction of cells, survival with complete elimination of viruses, or chronic infection with continuous coexistence of cells and viruses) crucially depend on mutation rates of the viral and Immunoglobulin proteins. Infection is always lethal if the virus mutation rate exceeds a certain threshold. Potent Immunoglobulins are discovered in this model via clonal selection and affinity maturation. Surviving cells acquire lasting immunity against subsequent infection by the same virus strain. As a second line of defense cells develop apoptosis-like behavior by reducing their lifetimes to eliminate viruses. These results demonstrate the feasibility of microscopic sequence-based models of immune system, where population dynamics of the evolving B-cells is explicitly tied to the molecular properties of their proteins.

**Introduction**

One of the most fascinating aspects of adaptive immunity is that it represents Darwinian evolution in action. Selection of the most fit B-cells expressing potent Immunoglobulins occurs through clonal selection and affinity maturation – processes that exemplify classic features of evolutionary selection (*1-3*). Enormous and fruitful experimental effort has been devoted to study specific mechanisms responsible for immune response, and the emerging detailed molecular picture confirmed earlier Burnet's hypothesis that some specificity for diverse antigens exists before these antigens are encountered (*3, 4*). A detailed micoscopic picture of maturation of affinity in immune response is emerging (*5, 6*).

Theoretical studies of immunity have been extensive and included analyses from population dynamics perspective (*7*) and in silico modeling of synapse formation and other molecular aspects of immune response (*8*). Phenomenological models of immune response have been reviewed in (*9*). Kamp and Bornholdt (*10*), Brumer and Shakhnovich (*11*) and Wang and Deem (*3*) considered the problem of co-evolution of host immune system (B-cells) and parasite (virus) in the quasispecies paradigm, assuming the existence of a master genome for both host and virus. Population dynamics approaches to study immunity have been successful in identifying global aspects of epidemics as well as optimal treatment strategies. A drawback of these approaches is in their phenomenological nature, which omits many microscopic details of the interaction between the host and parasite

Generally, a detailed mechanistic understanding of evolutionary processes is limited by lack of clear picture of genotype-phenotype relationship. In this regard immunity represents an important exception where direct relationship between antibody potency (a genotypic trait) and B-cell activation to proliferate (a phenotypic trait) has been found (*3*). Recently we developed a model that combines microscopic, sequence-based, description of dynamics of genes and stability of encoded proteins with cell population dynamics governed by exact relationship between genotype and phenotype (*12*). This model was applied in (*12*) to study early evolution of Protein Universe. Here we extend this line of modeling to study fundamental principles of immunity. We explicitly model interactions between viral and host defense (Immunoglobulin-like) proteins and assume that the strength of interaction between viral and host proteins affects the replication rate of both host cells and viruses, and therefore governs their population dynamics. In our model viruses infect Ig-producing cells, i.e. attack the immune system directly. As such this model is most directly applicable to class of infections that target immune system such as HIV or Epstein-Barr virus. We build the model bottom up based on simplest ''common-sense'' mechanistic assumptions and sequence-based molecular description of relevant proteins and their interactions. Specifically we make the following three mechanistic microscopic assumptions. *First,* we assume that stronger binding between immunoglobulin (Ig) proteins and viral antigens inhibits replication of viruses within the host cells (i.e. decreases viral replication rate). *Second,* to account for B-cell activation mechanism, we assume that the rate of B-cell division increases upon stronger binding between viral proteins and Ig-proteins. *Finally* we assume that the rate of B-cell division drops if Immunoglobulin proteins bind cell's own proteins, modeling the effect of autoimmunity. As mutations accumulate in both Immunoglobulins and viral proteins, selection pressure appears to optimize the Ig sequences for the strongest possible binding to the viral ones, whereas evolving viral sequences try to decrease the binding energy. This competition unfolds in the presence of the protein stability constraints, which limit the accessible repertoire of sequences. Indeed, strong binding between two sequences normally requires a hydrophobic patch on their surfaces; however, a hydrophobic surface may be unfavorable with respect to the folding free energy, normally decreasing protein stability (*13*).

Dynamics of the system establishes the qualitative outcome of an infection in this model: healing (survival of host cells with complete elimination of viruses), extinction of host cells, or development of a chronic infection, where the cells and viruses coexist, and the fraction of infected cells remains constant with time. Qualitatively, one can expect that for a very slowly mutating virus, cells will evolve a strongly virus-binding Ig protein, suppress virus replication within the infected cells, and eliminate the virus population. On the other hand, the speed at which such immunity evolves is fundamentally limited by the mutation rate of the Ig-like protein, and by the speed at which the favorable sequences spread through the population. Therefore, if the mutation rate of viruses is high enough, their sequence change can outpace the evolution of cellular response, resulting in a robust lethal infection.

. Below, we will present the results of simulations for this ab initio, microscopic model of immune response, and show that physics-based, sequence-level simulations can provide crucial insights, on all scales, into the development and dynamical regimes of immune response.

**Model**

In the model (see Supplementary Information for more details), each independent host cell carries one Ig-like protein and three other functional proteins. Proteins are modeled as 27-unit compact polymers, (*14*) so their thermodynamic properties, stability and interaction can be calculated exactly from their primary sequences (*12, 14*). Each virus carries one protein; once a virus infects a cell, the virus starts to replicate at a rate, $b^v$, dependent on the stability of viral protein and the interaction strength between viral and Ig proteins.

$$b^v = b_0^v \cdot P_{nat}(v) \cdot (1 - P_{int}(Ig, v)), \qquad (1)$$

where $b_0^v$ is a viral replication rate constant, $P_{nat}(v)$ is protein stability of virus (i.e. Boltzmann probability for the virus protein to be in its native conformation), and $P_{int}(Ig,v)$ is Boltzmann probability of a specific interaction between Ig protein and virus protein which serves as a measure of their interaction strength (see Supplementary Text for exact definition of this quantity) . Strong binding inhibits virus replication. Once the number of viruses in a given cell exceeds the lysis threshold, the viruses kill their hosts, are released, and can infect uninfected cells. If free viral particles cannot find any host cell, they are removed. The replication rate of the cells $b$ is determined by interaction between Ig and viral proteins if the cell is infected (strong binding increases cell replication rate), and the autoimmunity effect (strong binding between Ig and a functional protein decreases the replication rate),

$$b = b_0 \cdot \left[1 - \max_i P_{\text{int}}(Ig, g_i)\right] \cdot \min_j P_{\text{int}}(Ig, v_j), \tag{2}$$

where $P_{\text{int}}(Ig,g_i)$ and $P_{\text{int}}(Ig,v_j)$ are respectively the Ig protein's interaction strength with $i$-th normal functional protein of the cell and $j$-th viral particle in the cell, and indices $i$ and $j$ run from 1 to 3 and the number of viruses in the cell respectively. In the context of the immune system, an increase of cell replication rate with the antigen binding strength is the well-known phenomenon of B-cell activation by an antigen (*15-17*). Interestingly, the dependence of cell replication rate on the interaction strength is crucial for the development of strong immune response that results in complete elimination of the infection; without it strong immune response does not develop: all simulation runs converged to the chronic infection outcome (see Supplementary Information).

Upon division of an infected cell, viruses are randomly distributed between its two off-springs, and when an infected cell dies, any viruses it contains are also removed from the population. The simulation ensemble consists of up to 5000 completely independent cells and 20000 independent virus particles, mutating and replicating according to the above rules. If necessary, cell population is clipped at 5000, simulating a chemostat. The mutation rate of Ig proteins is set significantly above that of functional proteins, mimicking somatic hypermutation found in B-cells (*18, 19*), and the mutation rate of viruses is also higher than the normal host protein mutation rate. For the initial 1500 evolutionary time steps, cells are allowed to evolve with mutation rate $m=0.005$ per protein per time step to equilibrate their protein sequences, and then mutation rate drops to $m=0.0005$ (see Supplementary Information for details). At $t=2001$, one thousand of identical free viruses are introduced randomly into the system, starting the infection. In order to evaluate how mutation rates of Ig genes and viruses affect the outcome of cell-virus competition, we run multiple simulations with different mutation rates of viruses and Ig genes while keeping mutation rates of cell's genes fixed as described above.

## Results and discussion

Extensive simulations of the model over a broad range of parameters (B-cell and virus replication rates, and mutation rates of Ig and viruses) showed that the outcome of an individual simulation run falls into one of the three categories: healing, with complete extinction of the virus; extinction of the host (and the ensuing extinction of the virus); and chronic infection, where viruses and cells coexist with neither species winning the competition. The probability of each of the three outcomes depends on the parameters of the model, most notably on mutation rates of Ig and virus genes. Figures 1A and 1B present the population dynamics of cells and viruses, respectively, for representative simulation runs corresponding to the healing (H), chronic infection (C), and extinction (E) cases. One can see that in all cases, infection is followed by an incubation period during which the cell population remains constant, and the virus population increases until lysis threshold is achieved in the majority of cells. Ensuing lysis causes an abrupt drop of the cell population (Figure 1A). Duration of the incubation period depends on the virus replication rate. As the cell population decreases, cells that carry Ig proteins strongly binding to viruses acquire selective advantage, as their replication rate is higher, while virus replication within such cells is hindered. Accordingly, the population-averaged thermal probability of interaction between Ig and virus proteins $P_{int}$ (see Methods) sharply increases from a value of 0.25, characteristic for two random proteins under the simulation conditions, to about 0.7, see Figure 1C. This result points to the emergence of a strongly binding Ig protein as a result of population-level selection of most immune cells in the presence of a mutating virus and Immunoglobulin genes.

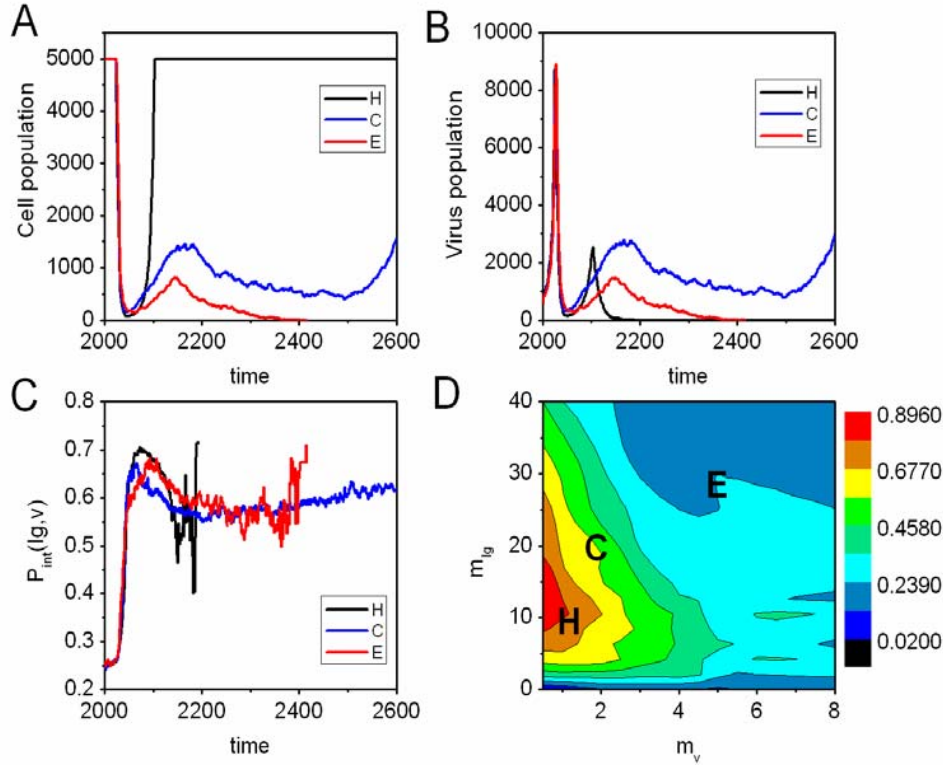

**Figure 1. Emergence of immunity.** *(A) Population of cells as a function of time in representative simulation runs resulting in healing (H), extinction (E) or chronic infection (C). Infection occurs at t=2001. (B) Population of viruses in the same simulation runs. (C) Averaged (over all Ig proteins and viruses) strength of interaction $P_{int}$ between viral and Ig proteins shows that strongly binding Ig's gain selective advantage, rapidly increasing $P_{int}$ as the cells try to overcome infection. (D) Phase diagram showing the probability of cell survival in 200 runs (color scale) as function of the mutation rates of Ig and viral proteins (relative to the mutation rates of functional proteins in the cell).*

Figure 1D represents a phase diagram of the model, where the color corresponds to the probability of cell survival at the given mutation rates of virus $m_v$ and of Ig protein $m_{Ig}$. The letters H, C, and E designate typical outcomes of the infections (healing, chronic infection, or host extinction) in the corresponding areas of the diagram. For a given combination of mutation rates, the outcome of an individual simulation run (representing an individual infection event) depends on the stochastic events leading to discovery or non-discovery of strongly binding Ig proteins and changes in stabilities of functional proteins (apoptotic pathway, see below). In many cases, the healing or extinction runs bear a significant similarity, and are different from the chronic infection pathway: healing and extinction pathways proceed through a narrow population bottleneck, whereas the population decrease in the chronic infection case is less drastic. Indeed, once the population had been sufficiently reduced, number fluctuations of both cells and viruses become significant, and may lead to the extinction of either viruses (in the healing case), or cells and viruses together in the extinction, or lethal infection case.

The microscopic modeling of interactions between cell and virus proteins in our model permits a detailed investigation of the emergence of immunity. The most direct path - hindering virus replication by evolving a strongly binding Ig protein - has been presented in Figure 1C. A more careful examination shows that the model cells also develop apoptosis as a complementary way of fighting the infection. Experimentally, apoptosis was found in naïve vertebrate hosts (*20-22*) as a first line mechanism of host defense. Plants, instead of recruiting B-, T-cells, and macrophages to eliminate the pathogens, activate the apoptotic pathway in the infected cells, in hope to eliminate the virus by killing its host cells thus stopping further virus production. Similarly, certain bacteria can induce premature activation of the lytic pathway of the infecting phage before virus assembly is complete, preventing the release of assembled viruses into the community (abortive phage resistance,(*23, 24*)). Interestingly, such a collective defense mechanism emerges naturally in our model. Indeed, in the model the cell lifetime is determined by the stability of the functional, non-Ig proteins. As seen in Figure 2B, after the infection, the life time of the cells decreases.

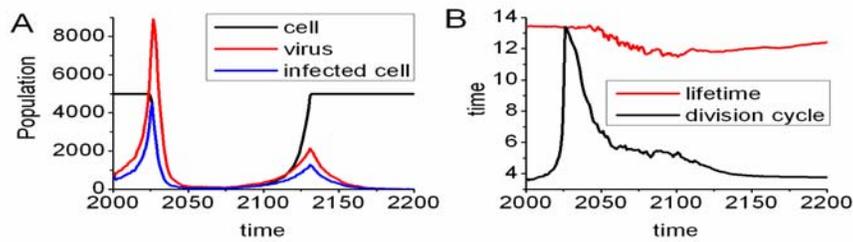

**Figure 2. The anatomy of a healing event. (A)** *Total population of cells, population of infected cells and viruses in a simulation run resulting in healing.* **(B)** *After the acute infection is over, average cell lifetime decreases through selection of cells with somewhat lower stability of their normal proteins, pointing to the emergence of apoptotic defense mechanism. The black line represents the average time between cell division events reflecting the effects of virus infection and subsequent activation.*

The apoptosis-like decrease of cell lifetimes is favorable for fighting infection, as viruses may not have enough time to reach the lysis threshold, if the cell terminates itself prematurely. In this way, viruses are removed from the population together with their host cells. After the infection had subsided, the life time of the cells gradually returns to pre-infection levels (Figure 2B). By comparing Figures 1C and 2B, we conclude that in the model, the rapid response to the infection is evolution of strongly binding Ig sequences, while the decrease of cell lifetime via lowered stability of non-Ig proteins occurs at longer times when the acute phase is over. Interestingly, CD4+ T cell apoptosis has been found as one of the responses to HIV infection (*2, 3*).

Selection of specific, strongly binding Ig proteins from a vast pool of initial prototypical B-cells carrying a very large set of un-optimized Ig sequences proceeds in our model via clonal selection with subsequent affinity maturation. In Figure 3A we present the histogram of the Ig-to-virus binding strength $P_{int}$ distributions at four time slices. Immediately after infection, $t$=2001 (black curve), most of the Ig sequences exhibit a marginal binding to the viral antigen with $P_{int}$~0.25, typical for two random proteins in this model. However, a small fraction of B-cells carries Ig molecules with a relatively strong binding, $P_{int}$>0.4 (tail of the distribution, shaded area). Subsequently, such cells enjoy selective advantage (see Eq. (2)), and their fraction in the population increases ($t$=2020, red curve). This is similar to the natural clonal selection process.(*17*) Furthermore, as the (moderately) binding sequences have been discovered, they undergo mutations with further selection, corresponding to affinity maturation increasing the binding strength $P_{int}$ to its final value of ~0.7 ($t$>2050, green and blue curves),. To confirm that the strongly binding Ig's are normally direct descendants of the original moderately binding sequences selected at the first – clonal selection – stage of the immune response, we marked cells with $P_{int}$>0.4 as ''red'' when infection first occurred and their ancestors have never been infected, and postulated that their progeny retains the ''red'' color. In Figure 3B, we plotted the fraction of the ''red'' cells in the population. One can see that the fraction of the ''red'' cells initially increases with time, and reaches its maximum as the cells recover from the infection (cf. Figure 1). Therefore, virus removal is mostly accomplished by the direct descendants of the few Ig sequences that accidentally had a certain binding affinity and have been amplified and optimized via a two-stage process of clonal selection and affinity maturation. Our simulations suggest that the primary role of clonal selection is to amplify B-cells with moderately binding sequences from a diverse pool that evolved before the antigen was presented. As a next step, cells carrying clonally selected Ig genes undergo improvement by specific mutations dramatically increasing the binding strength. Evidently, such mutations can be efficiently discovered only during mutagenesis in a pool of moderate binders, rather than by pure chance among pseudorandom sequences. Importantly the fraction of progenitors of clonally selected ''red'' cells drops after infection is eliminated but it returns to the level which is higher than initial one (see long-time plateau in Fig.3B) reflecting effect of memory. Indeed, upon re-infection the cells exhibit memory effect in their immune response (data not shown).

Next, we study the distribution of sequences of evolving Ig proteins. In Figure 3C, we present the measure of sequence diversity - sequence entropy - of evolved Ig sequences (black curve) and Ig-to-virus binding strength $P_{int}$ (red curve) as function of time. The figure shows that discovery of strongly binding Ig proteins, manifested in a sharp increase in $P_{int}$, is accompanied by a drop in the sequence entropy of Ig proteins. As the infection is being eliminated, the Ig proteins have the minimum sequence entropy and are essentially monoclonal, similar to the production of large quantities of specific, monoclonal B-cells in vertebrate immune systems upon infection.

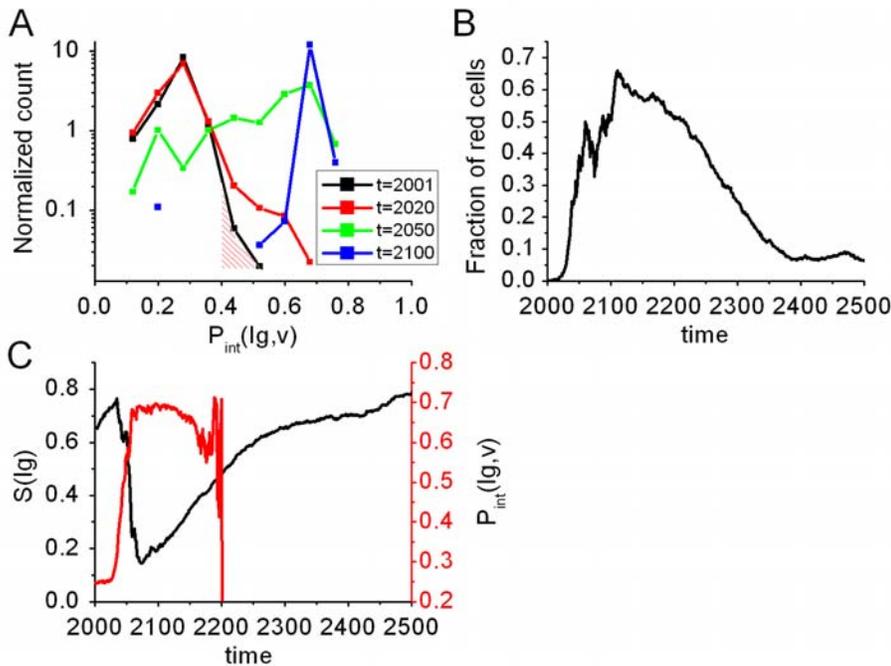

**Figure 3.**

**Clonal selection and affinity maturation.** *(A) Histogram of the binding strength $P_{int}$ in the ensemble of Ig molecules at various times. From t=2001 to t=2020, clonal selection increases the fraction of Ig's strongly binding the virus: the tail of the distribution becomes more prominent, while the maximum is unmoved. As time goes on, mutations in the previously selected Ig genes further increase $P_{int}$ and shift the distribution to the right until monoclonal population of strongly binding Ig molecules emerges. (B) We marked the cells with $P_{int}>0.4$ at t=2001 (shaded area in (A)) as ''red'' and followed the fraction of their progeny. As the infection is being removed, most of the Ig's are descendants of the few strongly binding sequences amplified via clonal selection. (C) Sequence entropy of the evolving Ig proteins (black line) and their binding strength $P_{int}$ as function of time (ref). Development of immunity is accompanied by a rapid increase of $P_{int}$ and a decrease of the sequence entropy of Ig proteins, confirming the appearance of a monoclonal Ig population in response to infection.*

The two-stage process of discovery of potent Ig proteins explains existence of optimal rate of somatic hypermutation for Ig genes at which immune response is most efficient (see Fig.1D). Qualitatively, this is due to the fact that affinity maturation occurs in population of already pre-selected (via clonal selection) cells carrying moderately potent Ig proteins. In this case too low mutation rate of Ig proteins will prevent affinity maturation while too high mutation rate could result in predominantly deleterious mutations which destroy even moderate binding of Ig to antigens developed at the clonal selection stage. A more quantitative analysis provided in Supplementary Text suggests that optimal mutation rate *per amino acid* in Ig is determined by the relation:

$$m^*t = -\frac{\log\left(1-e^{-\frac{(\Delta G_0)^2}{2\sigma^2}}\right)}{e^{-\frac{(\Delta G_0)^2}{2\sigma^2}}} \quad , \tag{3}$$

where $t$ is a characteristic time of affinity maturation (which may depend on replication rate of the virus), $\Delta G_0$ is average change of binding free energy per amino acid upon mutation in the population of clonally selected Ig proteins, and $\sigma$ is standard deviation of this quantity in a statistical pool of mutations. The result given by Eq.3 clearly highlights the role of clonal selection in determining the optimal Ig mutation rate. Indeed in the ensemble of Ig molecules that had been initially selected for antigen binding, random mutations have in average a detrimental effect on binding energy. Hence, the value of $\Delta G_0$, the energetic effect of a random mutation on the antigen affinity of binding Ig molecules will be significant. Stronger clonal selection at the first stage leads to increase of $\Delta G_0$, and the optimal rate approaches its absolute limit of just one mutation per variable part of Ig in affinity maturation process. Alternatively, in unselected naïve population, $\Delta G_0 = 0$, which means that a mutation has equal probability to strengthen and to weaken binding to the antigen. In this case there is no optimal rate of Ig mutations – the faster the better. Interestingly the optimal mutation rate observed in our simulations – roughly two amino acid mutations per variable region in affinity maturation process

(see Supplementary Text and Figure S2) – is very close to the hypermutation rates (HMR) observed in Nature (*17*). Further our theory predicts that optimal HMR does not depend on virus mutation rate – in agreement with simulations (see Fig.1D) and estimates that show similar HMR for broad range of infections (*11, 25, 26*).

**Conclusions**
We presented a simple, microscopic model of immunity in an interacting host-pathogen system. We built the model entirely bottom-up, starting from protein sequences, calculating their thermodynamic properties, such as stability and ability to interact with each other, and making simple, well-defined assumptions about the effects of protein stability and interactions on the reproductive rates of both cells and viruses. A complex interplay between genetics, physics of proteins, and population dynamics provides a rich variety of dynamical regimes, identifying all possible outcomes such as healing, host extinction, and chronic infection as function of the mutation and replication rates. We have demonstrated that these minimal assumptions are sufficient for the discovery of monoclonal Ig-like virus-binding proteins in a two-stage process that involves clonal selection and affinity maturation, and, more surprisingly, for uncovering the apoptotic pathway as a means of collective defense against the infection. Microscopic consideration of the sequences allowed a clear dissection of the clonal selection and affinity maturation mechanisms, which allow for fast and efficient selection of specific, strongly binding Ig sequences from an initial pool of randomized, non-specific Immunoglobulins. We have shown that B-cell activation followed by fast clonal selection is indispensable for overcoming the infection, as a direct discovery of Ig sequences from a random pool is prohibitively inefficient and persistently leads to chronic infection. The presented diagram of the infection outcomes as function of the mutation rates can be of interest for the development of therapeutic approaches to certain diseases.

Our model reproduces qualitatively all phases of the dynamics of HIV infection – acute first viremia, initial rapid clearance due to emerging immune response, nadir, latent phase and final viremia at the onset of AIDS (see blue chronic infection curve on Fig.1B).

A population dynamics view on immunity emphasizes competition between host and pathogen (*9*). The present microscopic model, while confirming this aspect of immune response, points out to an important aspect of immunity which is independent on dynamics of viral mutational load. Indeed, two-stage process of selection of potent Immunoglobulins requires mutation dynamics that is to a large extent independent on the mutation rates of viruses – as can be seen on the phase diagram Fig.1D which shows that no healing can occur when mutation rate of Immunoglobulins is too low, regardless of viral mutation rate. However, necessary maturation time (and therefore optimal SHM rate in Eq.(3)) may depend on dynamics of viral *replication.* This finding points out to the necessary and universal character of SHM.

An important aspect of our microscopic model is autoimminity which significantly restricts sequence evolution of Immunoglobulins. Deem and coworkers suggested that cross-reactivity may affect the affinity maturation dynamics (*27*). Further analysis of our model can shed more light on the interplay between autoimmunity and dynamics of sequence selection of potent Immunoglobulins.

One important observation in our model is that healing may occur only via a crucial population bottleneck when the number of cells drops dramatically. Such scenario is perhaps not realistic for immune infections, like HIV (where T-cell count is essential for organism's survival) and Epstein-Barr virus (infecting B-cells). Therefore most likely outcome of an infection targeting immune system is a chronic pathway and vaccination may be not efficient to fight this outcome. However, therapies that provide additional elimination of viruses by lowering their replication rates may help to avoid the bottleneck scenario and result in clearance of viral titer in the population.

Our model is still quite minimalistic as it uncovers general robust features of immune response on various scales while exact molecular mechanisms here are quite schematic and differ significantly from actual biological pathways that operate in living cells. Nevertheless the striking similarity of the mechanism of adaptive immunity which are found here to those discovered in jawed vertebrates attests to the robustness and certain degree of inevitability of evolutionary processes that lead to adaptive immunity. It is quite remarkable to observe the anatomy of evolutionary events - how complex biological functions and responses emerge as a result of mutation and natural selection processes as postulated by Darwin long ago.

*Acknowledgements.* We thank Michael Deem for helpful comments on the manuscript.

## Supplementary Text

## Methods

**Protein model: Stability and interactions**

We consider a 3x3x3 lattice protein as a model protein in order to calculate thermodynamic quantities such as protein stability and protein interaction strength exactly (*1, 2*). To save computer time, we reduce protein structural space from 103,346 structures to randomly and uniformly chosen 10,000 structures. Each protein folds into the native structure which has minimum energy out of 10,000 structures. Protein stability is calculated in terms of the Boltzmann probability ($P_{nat}$) of the native structure of the protein at given temperature,

$$P_{nat}(g,T) = \frac{\exp[-E_0(g)/T]}{\sum_{i=1}^{10000} \exp[-E_i(g)/T]}, \quad (S1)$$

where $g$ is a protein sequence, $E_0(g)$ and $E_i(g)$ are energies of protein in the native and $i$-th structure respectively, and $T$ is an environmental temperature. In order to calculate interaction strength between two proteins, we consider rigid-body docking between two 3x3x3 lattice proteins. Each lattice protein has six binding surfaces and four-fold rotational symmetry of a binding surface (only the strongest binding positions sharing 9 interaction bonds are taken into account). Therefore, a pair of lattice proteins has 6x6x4 binding conformations. Protein interaction strength is defined by the Boltzmann probability ($P_{int}$) of the native binding between two protein $g_1$ and $g_2$ as following.

$$P_{int}(g_1,g_2) = \frac{\exp[-f \cdot E_0(g_1,g_2)/T]}{\sum_{j=1}^{144} \exp[-f \cdot E_j(g_1,g_2)/T]}, \quad (S2)$$

where $E_0(g_1, g_2)$ is the minimum binding energy of two proteins out of 144 binding modes, and $f$ is a pre-factor that takes into account possible different strengths of intra-protein and inter-protein interactions. We use the Miyazawa-Jernigan pairwise contact potential for both protein structural and interaction energies (*3*), but scale protein-protein interactions by a constant factor. We chose $T = 0.85$ in Miyazawa-Jernigan dimensionless energy unit and $f = 1.2$.

**Cell dynamics; reproduction, death, immunity, and autoimmunity**

We develop a simple model cell whose life is described by genetic information encoded in its genome. The genome of a cell consists of four genes. The first gene, named Ig (immunoglobulin-like) gene, is assigned for the function of immunity, and the remaining three genes are normal functional genes which control the life of the cell. In the model, Ig protein's interaction strength is responsible for the molecular recognition and suppression of extrinsic proteins. Once a cell is infected by a virus, it can obtain immunity against the virus by increasing the Ig protein's interaction strength to viral RNA or enzymes. Also, the Ig protein should recognize "self" genes from "non-self" genes. If it mistakes "self" proteins for "non-self" proteins and binds them strongly, it suppresses the function of cellular proteins. This is the main cause of autoimmunity and it is incorporated into the reduction of cell division rate $b$ which describes as follows

$$b = b_0 \cdot \left[1 - \max_i P_{int}(Ig, g_i)\right], \tag{S3}$$

where $b_0$ is a cell division rate constant and $P_{int}(Ig,g_i)$ is the interaction strength between Ig and $i$-th normal functional protein in the cell and index $i$ runs from 1 to 3. A loss of a functional protein's stability results in malfunction of life controlling system of the cell (*1*). We connect the minimum stability of any functional protein in a cell to death rate $d$ of the cell as following.

$$d = d_0 \left[1 - \min_i P_{nat}(g_i)\right], \tag{S4}$$

where $d_0$ is a cell death rate constant and $P_{nat}(g_i)$ is $g_i$ protein's stability and index $i$ runs from 1 to 3. As the population grows, cells with lower death rate to spread out their offspring due to longevity. This selection pressure enriches the system with extremely stable functional protein as a majority. In order to avoid the system being driven to the extreme, we introduce the concept of marginal stability in the model. The upper limit of $P^*_{nat}$ is set in the model and $P_{nat}$ value is substituted with the upper limit when it exceeds the upper limit, so the there is no pressure towards evolving proteins with native state stability above $P^*_{nat}$.

**Virus dynamics; replication and lysis**

A viral particle carries a single protein (antigen), denoted by $v$. A free viral particle can infect a cell which is unoccupied by other viral particles with viral infection rate. This particle can be mutated or replicated only when it resides in an infected host cell. When the population of the viral particles in a living host cell exceeds the lysis threshold, the viral particles bring about lysis in the cell and are released. If free viral particles cannot find an unoccupied (uninfected) host cell, they are removed. The replication rate of virus is regulated by its protein stability and protein interaction strength with Ig protein. A viral replication rate $b^v$ can be written as following.

$$b^v = b_0^v \cdot P_{nat}(v) \cdot \left(1 - P_{int}(Ig, v)\right), \tag{S5}$$

where $b_0^v$ is a viral replication rate constant, $P_{nat}(v)$ is protein stability of virus, and $P_{int}(Ig,v)$ is protein interaction strength between Ig and viral protein. Once a virus infects a cell, it takes over the resources of the cells and reduces the resources available for cell reproduction. Therefore, the reproduction rate of infected cell is also regulated by Ig protein's interaction strength with viral protein. Viral infection modifies cell reproduction rate as follows:

$$b = b_0 \cdot \left[1 - \max_i P_{int}(Ig, g_i)\right] \cdot \min_j P_{int}(Ig, v_j), \tag{S6}$$

where $P_{int}(Ig, v_j)$ is Ig protein's interaction strength with $j$-th viral particle in the cell, and index $j$ runs 1 to the number of viruses in the cell.

## Simulations

Initially 100 cells are seeded with 4 identical primordial genes. Initial simulation runs with cell reproduction rate constant, $b_0 = 0.339$ and cell death rate constant, $d_0 = 0.145$. Due to the differences between the time scales of protein evolution and a much faster response to a viral infection, the population dynamic simulation proceeds in two steps. For 1500 initial time steps, cells are allowed to evolve with mutation rate of functional proteins in the cell, $m=0.005$ (*attempted* mutations per gene per unit time) in order to equilibrate the system. Then, the *attempted* mutation rate is reduced to $m=0.0005$ and the cells evolve for 500 time steps to re-equilibrate system at the new mutation rate. At $t=2001$, 1000 identical free viruses are introduced in the system, starting the infection with the viral infection rate, 0.75. The sequence of viral protein is randomly chosen before simulation, whose initial stability, $P_{nat}=0.5$. The upper limit of $P_{nat}$ for functional and viral protein is set to 0.5. The viral replication rate constant $b_0^v$ and the lysis threshold are set to 2.27 and 4. After infection, the system evolves up to $t=4000$.

## Sequence entropy calculation

In order to observe selection of monoclonal Ig proteins, we calculated the sequence entropy of Ig proteins. The sequence entropy of a residue in $k$-position is defined as following (*4*).

$$S_k = -\sum_{i=1}^{20} P_i^k \log P_i^k \tag{S7}$$

, where $P_i^k$ is a probability to find $i$-type amino acid in $k$-position. The sequence entropy is then averaged for all 27 positions in a sequence.

## The importance of cell activation by antigen – cell reproduction rate dependence on Ig-virus interaction strength.

Viral infection interferes with cellular functions related to cell cycle(*5*) and, most importantly recognition of antigen of B cells stimulates their proliferation. In our model, we incorporate this effect into Eq. (6). In order to study the importance of the relation between viral infection and cell reproduction rate, we ran control simulations where we neglected the dependence on Ig-virus interaction strength, $P_{int}(Ig, v)$, i.e. for this control simulation we modified the Eq (6) as follows:

$$b = b_0 \cdot \left[1 - \max_i P_{int}(Ig, g_i)\right], \tag{S8}$$

We ran 200 control simulations with modified cell reproduction rate dependence Eq.(S8). No healing or death outcomes were observed, because viral infection without interfering cell cycle does no harm to cells. Population of cells still keeps growing in Figure S1A and S1D. Population of viruses exhibits exponential growth up to t=2055, then fluctuates around 9,000 because there exists no uninfected cell any more (Figure S1B). In Figure S1C, $P_{int}(Ig,v)$ cannot reach as the high level as in original simulation, because no decrease of cell population weakens evolutional pressure to select cells with higher immunity. Reproduction of cells in control simulations is not affected by viral infection (Figure S1D), and no apoptosis is induced.

## Derivation of the optimal Ig mutation rate

Consider affinity maturation time scale $t$. At this time scale $N$ mutations can occur and we are interested in the probability that affinity will be improved. Since we mutate Ig's which represent a biased (towards stronger binding) ensemble that was are already pre-selected (via clonal selection), then a mutation is statistically more likely to be destabilizing. Assume Gaussian probability density distribution for the mutation effect on binding free energy (6):

$$P(\Delta G) = \frac{1}{\sqrt{2\pi\sigma^2}} \exp\left(-\frac{(\Delta G - \Delta G_0)^2}{2\sigma^2}\right) \quad (S9)$$

Here $\Delta G_0$ represents the average effect of each mutation. $\Delta G_0 > 0$ means that mutations statistically tend to be destabilizing; this is the case for the preselected ensemble of Ig molecules biased towards stronger binding to an antigen. $\sigma$ is standard deviation of the energetic effect of single mutations. The effect of $N$ mutations occurring in a replication cycle is additive, hence probability density should scale according to Central Limit Theorem (assuming that the mutations are uncorrelated and independent):

$$P_N(\Delta G) = \frac{1}{\sqrt{2\pi N\sigma^2}} \exp\left(-\frac{(\Delta G - N\Delta G_0)^2}{2N\sigma^2}\right) \quad (S10)$$

The probability that $N$ mutations occur in the given time frame $t$ follows the Poisson distribution:

$$I_N = \frac{(mt)^N e^{-mt}}{N!} \quad (S11)$$

where $m$ is *amino acid* mutation rate *per variable Ig domain*. The probability that mutations improve binding, i.e. that affinity maturation could occur, is:

$$P_{AM} = \int_{-\infty}^{0} d(\Delta G) \sum_{N=1}^{\infty} I_N P_N(\Delta G) \quad (S12)$$

$$\int_{-\infty}^{0} d(\Delta G) P_N(\Delta G) = \frac{1}{\sqrt{2\pi N \sigma^2}} \int_{-\infty}^{0} d(\Delta G) \exp\left(-\frac{(\Delta G - N\Delta G_0)^2}{2N\sigma^2}\right) = \frac{1}{\sqrt{\pi}} \int_{-\infty}^{-\sqrt{N}\frac{G_0}{\sqrt{2}\sigma}} \exp(-q^2) dq,$$

where we used the change of variables:

$$q = \frac{\Delta G - N\Delta G_0}{\sqrt{2N\sigma^2}}$$

The last integral is error function which can be asymptotically approximated as

$$\int_{-\infty}^{0} d(\Delta G) P_N(\Delta G) \approx \frac{1}{\sqrt{\pi}} \exp\left(-N\frac{(\Delta G_0)^2}{2\sigma^2}\right)$$

Finally we get:

$$P_{AM} = \frac{1}{\sqrt{\pi}} \sum_{N=1}^{\infty} \frac{(mt)^N}{N!} e^{-mt} e^{-N\frac{(\Delta G_0)^2}{2\sigma^2}} = \frac{1}{\sqrt{\pi}} \exp\left(mt\left(e^{-N\frac{(\Delta G_0)^2}{2\sigma^2}} - 1\right)\right) - \exp(-mt) \quad (S13)$$

$P_{AM}$ has a maximum at

$$m^* t = -\frac{\log\left(1 - e^{-N\frac{(\Delta G_0)^2}{2\sigma^2}}\right)}{e^{-N\frac{(\Delta G_0)^2}{2\sigma^2}}} \quad (S14)$$

which is the Eq.(3) of the main text.

### Comparison of the optimal model Ig mutation rates with experiment.

The *attempted mutation* rate of the ''normal genes'' in the model is $m_{normal} = 5 \cdot 10^{-4}$ per time step per gene. The optimal attempted mutation rate for Ig-genes is roughly 10 times greater (see Fig.1C), i.e. it is $m_{Ig}^{opt} = 5 \cdot 10^{-3}$ per gene per time step. However, importantly, the actual rate at which amino acid mutations accrue in the population of Ig genes is higher – it is about $2 \cdot 10^{-2}$ per time step per gene (see Fig.S2). The difference is due to the fact that Ig genes are under selection and populations get enriched with Ig's that accumulated favorable mutations. Similar effects contribute to the overall estimate of the SHM rates in real B-cells (*7*). The estimated SHM rates is $10^{-3}$ per bp per generation (*7, 8*) which translates roughly into 0.3 mutations per variable Ig gene (110 amino acids) per generation. In our simulations cells divide roughly once in 10 time steps, i.e. in our case we effectively observe 0.2 mutations per Ig gene per generation at optimal Ig mutation attempt rate. This is close to the experimental value of 0.3, and the difference may be due to different relative sizes of antigen recognition sites in our small model proteins and in larger real Immunoglobulins (apart from the fact that in real variable parts of Ig genes mutation rates may be higher in hot spots such as CDRs (*8, 9*) . Further, in our case full affinity maturation occurs in 100 time steps – roughly 10 generations ( see Fig. 3A) which is again close to real situation (*7*).

# References for the Supplementary Text

# Supplementary Figures

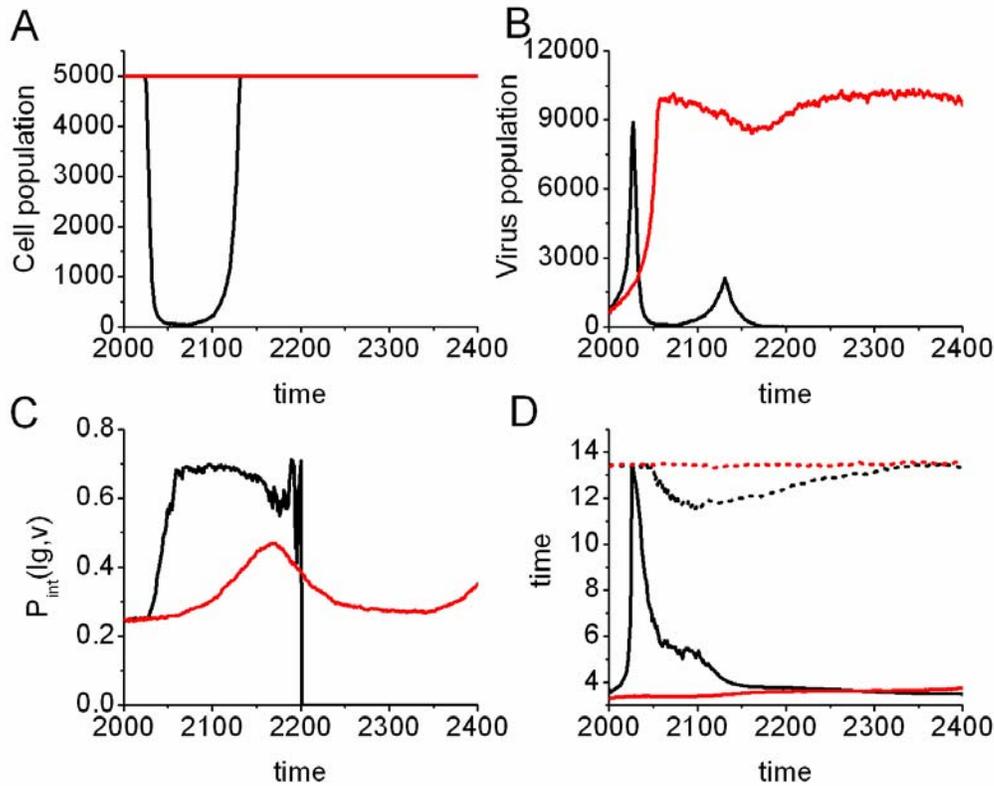

Figure S1.

**Control simulation without cell activation by antigen i.e. without cell replication rate dependence on $P_{int}(Ig,v)$ (red curves).** *For comparison purposes we also provide here the results from the healing pathway in the original simulations where cell activation is included (black curves) (A) Population of cells does not decrease without cell activation. (B) Population of viruses saturated around 9,000 and no healing occurs in control simulations. (C) No selection for strongly binding Immunoglobulins: Ig-virus interaction strength in control simulation is significantly lower than one in original simulation. (D) The solid and dotted lines, respectively, represent averaged cell division time (1/b) and lifetime (1/d) as function of simulation time. No dramatic change in cell division time and no apoptosis is observed in control simulation.*

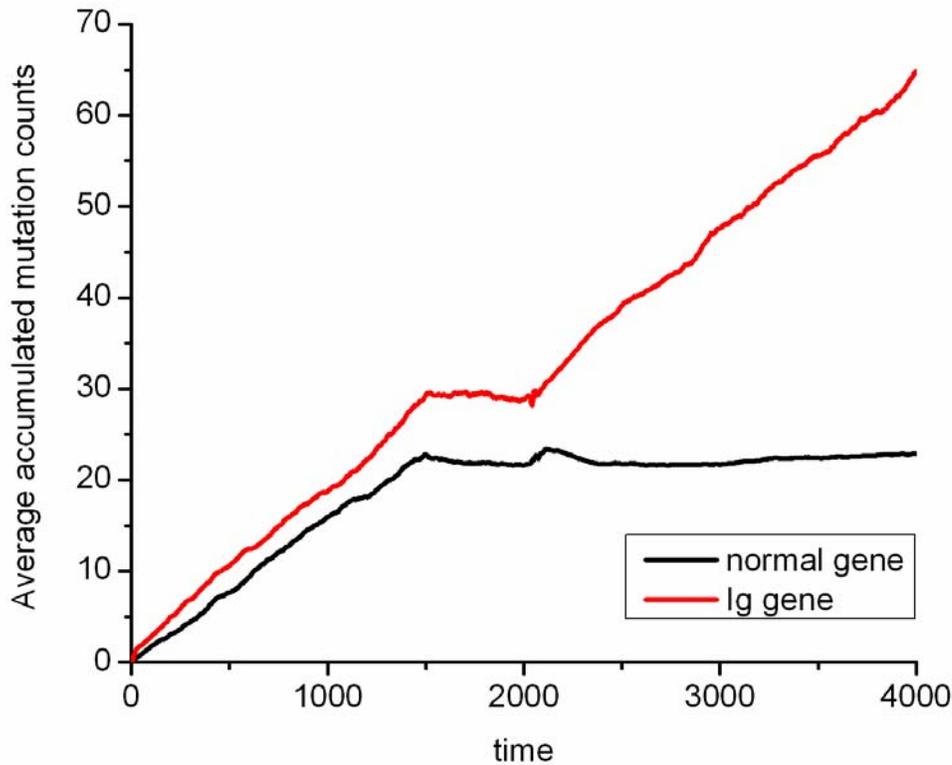

Figure S2.
**Average accumulated mutation counts of normal functional and Ig gene of cells.** *The black and red lines respectively show the averaged effective mutation counts of normal and Ig gene which are accumulated in the genome of survived cells during simulation. This quantity is calculated by tracing ancestry of each gene to the original progenitor gene and averaging over all exiting genes in the population at each time point. Note that this quantity is different from attempted mutation rates due to effect of selection on both ''normal genes'' of the cell and Immunoglobulin genes. Before t=1501, we observed different accumulation mutation rates of 0.0153 and 0.0179 (mutations per gene per unit time) for normal genes and Ig genes (reflecting higher selection pressure on Ig due to autoimmunity constraint), even though we apply the same attempted mutation rate, 0.005 to both genes. At t=1501 the attempted mutation rate drops ten-fold for both normal and Ig genes. Infection occurs at t=2001 at which time attempted mutation rate for Ig genes increases in $m_{Ig}$ times (10 in this example corresponding to optimal mutation rates for Ig in Fig 1D). High positive selection pressure is exerted on Ig gene after infection resulting in higher rate of accumulated mutations ($2 \cdot 10^{-2}$) than the rate of attempted Ig mutations ($5 \cdot 10^{-3}$). After virus infection at t=2001, normal genes accumulate some mutations in order to induce apoptosis and then dissipate the accumulated mutations after healing. This results in apparent negative slope of the curve for normal genes (black) which*

*reflects population effect of post-healing survival of cells that accumulated smaller number of mutations and had more stable proteins.*